\documentclass[longbibliography,twocolumn,showpacs,preprintnumbers,floatfix,amsmath,amssymb,superscriptaddress,pre]{revtex4-1}
\usepackage[english]{babel}
\usepackage{graphicx}    
\usepackage{bm}          
\usepackage{braket}
\usepackage[usenames,dvipsnames]{xcolor}
\usepackage[colorlinks=true ,citecolor=blue]{hyperref}

\newcommand{\comment}[1]{}
\newcommand{\ee}{\mathrm{e}}
\newcommand{\ii}{\mathrm{i}}

\begin{document}
\title{Solitons in one-dimensional lattices with a flat band}
\author{Dario Bercioux}
\email{dario.bercioux@dipc.org}
\affiliation{Donostia International Physics Center (DIPC), Manuel de Lardizbal 4, E-20018 San Sebasti\'an, Spain}
\affiliation{IKERBASQUE, Basque Foundation for Science, Maria Diaz de Haro 3, 48013 Bilbao, Spain}
\author{Omjyoti Dutta}
\affiliation{Donostia International Physics Center (DIPC), Manuel de Lardizbal 4, E-20018 San Sebasti\'an, Spain}
\author{Enrique Rico}
\affiliation{IKERBASQUE, Basque Foundation for Science, Maria Diaz de Haro 3, 48013 Bilbao, Spain}
\affiliation{Department of Physical Chemistry, University of the Basque Country UPV/EHU, Apartado 644, E-48080 Bilbao, Spain}

\begin{abstract}
We investigate the spectral properties of a quasi-one-dimensional lattice in two possible dimerisation configurations. Both configurations are characterised by the same lattice topology and the identical spectra containing a flat band at zero energy.  We find that, one of the dimerised configuration has similar symmetry to a one-dimensional chain proposed by Su-Schrieffer-Heeger for studying solitons in conjugated polymers. Whereas, the other dimerised configuration only shows non-trivial topological properties in the presence of chiral-symmetry breaking adiabatic pumping. 
\end{abstract}

\maketitle

\section{Introduction}
Lattice systems with flat bands have been of great interest in the past years. On one side, research was devoted in finding possible correlated states leading to fractional quantum Hall states without an applied magnetic field~\cite{Tang:2011,Neupert:2011,Sun:2011}. In parallel, there has been a research activity devoted to investigate the presence of flat bands in two-dimensional (2D) lattices as the $\mathcal{T}_3$~\cite{Bercioux:2009,Bercioux:2011,Dutta:2015} or the Lieb~\cite{Goldman:2011} lattice for possible application to cold atom physics in optical lattices~\cite{Moller:2012,Jo:2012,Aidelsburger2014,Dutta:2015b}, photonic crystals~\cite{Vicencio:2015,Mukherjee:2015} and exciton polaritons~\cite{Schmidt:2016,Baboux:2016}. These 2D systems~\cite{Lin:2015} --- both characterised by a pseudo-spin 1 --- present peculiar transport properties,  as the so called \emph{super Klein tunnelling}, where the tunnelling probability for a particle going through a potential barrier is unitary independently of the incidence angle and barrier length~\cite{Shen:2010,Urban:2011,Dora:2011,Lan:2011}. This scattering mechanisms has also potential applications in the framework of photonic crystals~\cite{Fang:2016} and quantum optics~\cite{Leykam:2016}. Several other pioneering works have addressed the role of the flat band in photonic systems~\cite{Nakata:2012,Masumoto:2012,Mukherjee:2015} and cold-atom gases~\cite{Jo:2012,Aidelsburger2014}, especially the key role played by geometric frustration was highlighted~\cite{Nixon:2013} and some characteristic features, such as the absence of transverse diffraction of wave-packet in a flat band~\cite{Mukherjee:2015}. It should be noted that, one can obtain such  in the propagation of photonic wave-packet due to Anderson localisation~\cite{Segev:2013}.  

In the past few years, the attention has moved to search of three-dimensional (3D) structure that presents a Dirac cone with a flat band in the middle of the cone, this corresponds to a 3D representation of pseudo-spin~1. In a first proposal based on HgCdTe quantum well, this triple degeneracy point has been achieved by playing with doping level of Cd in the quantum well heterostructure. This triple degeneracy point has also been observed experimentally~\cite{Orlita:2014,Malcolm:2015,Teppe:2016}. This point is however accidental and not robust against external perturbation. In a 2016 theoretical work, Bradlyn \emph{et al.}~\cite{Bradlyn:2016} use irreducible representation of the little group of lattice symmetries to show that there exists space groups that hosts naturally these type of triple degeneracy points. As degeneracy points are enforced by lattice symmetry, these are intrinsically protected against external perturbations. Their prediction is further confirmed by \emph{ab initio} calculations. 

In the present work we investigate topological property of a dimerised one-dimensional chain containing a flat band. Our prototype model is the one-dimensional dimer chain, or one-dimensional Su-Schrieffer-Heeger (SSH) model, originally introduced to describe solitonic effects in polymers~\cite{Su:1979,Su:1980,Heeger:1988}. Such a dimer chain supports edge states when it is in the topologically non-trivial phase~\cite{Asboth:2016}. The SSH model shows topological non trivial properties also in the presence of Coulomb interaction~\cite{Bello:2016}. 

\begin{figure*}
\begin{center}
\includegraphics*[width=\textwidth]{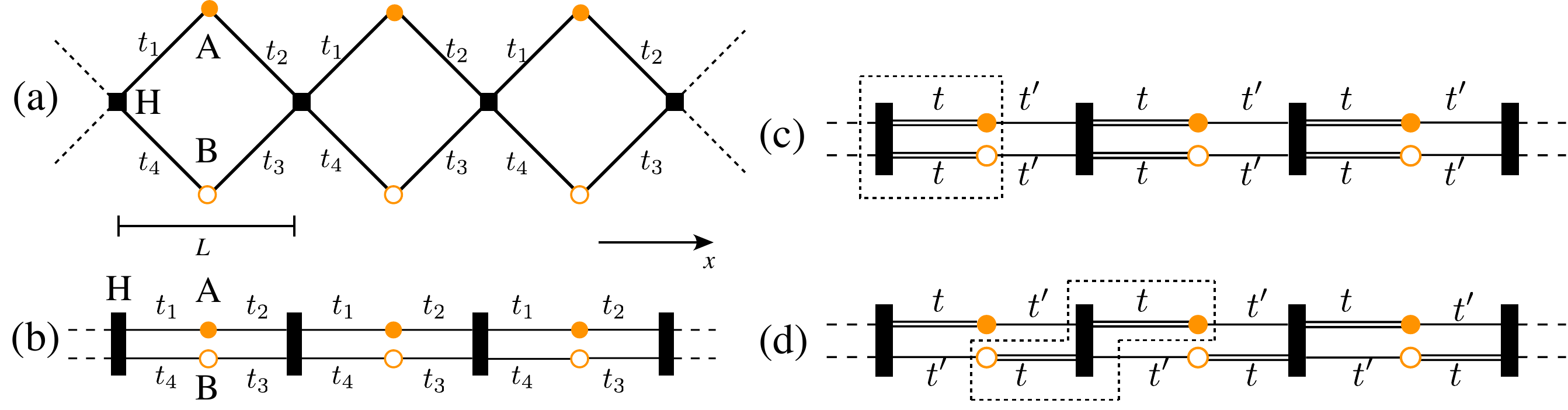}
\caption{\label{system} Sketch of the diamond chain (a); same lattice but stretch in a quasi-one-dimension form (b); dimerised form of the diamond chain: lattice I (c); lattice II (d), the dashed boxes show the two unit cells. The two lattices differs for the position of the double bonds, in lattice I these are parallel whereas in lattice II they are shifted of a lattice site.}
\end{center}
\end{figure*}

In a  quasi-one-dimensional lattice, the flat band can be obtained by considering a bipartite lattice structure with a unit cell containing three lattice sites with unequal connectivity [Fig.~\ref{system}(a)] --- we name this lattice in the following diamond chain~\cite{Bercioux:2004}. In this system, there is a lattice site ---  named Hub (H)  --- that has four next-neighbour, and two Rim sites, A and B, respectively. These have only two next neighbour that are always H sites; it is not possible in this system to move directly from A to B, this is allowed only via an H site. This bipartite structure with  inequivalent sites in the unit cell is at the origin of the flat band~\cite{Bercioux:2009,Lin:2015}. The interest in investigating this diamond lattice roots in the possibility to implement two distinct  dimerization of the lattice.  We will illustrate --- via fundamental symmetry arguments --- that the two dimerisation configurations lead to opposite topological properties.
A structure as the diamond chain is also interesting because it has been proposed recently an implementation with optical lattices by Hyrkas \emph{et al.}~\cite{Hyrkas:2013} This is not the only possibility to implement a lattice chain with a flat band, other lattices with flat bands are present in the literature~\cite{Huber:2010,Schulze:2013}. 
%
%
%
%
This article is organised in the following way: in  Sec.~\ref{ModForm} we present the  diamond chain in the two possible dimerisation configuration we have disclosed. In Sec.~\ref{topo} we present the topological properties of these two lattice configurations. We conclude the manuscript with a discussion of our finding and  suggestions for possible experimental implementations.

\section{Models and Formalism}\label{ModForm}

We consider a prototypical quasi-one dimensional lattice with a flat band: a diamond chain containing three distinct sites in the unit cell [see Fig.~\ref{system}(a)]. In Fig.~\ref{system}(a) we denote the three sub-lattices as H, with coordination number four, Rim A and B (or simply A and B), both with coordination number two. The drawing with a square path as in Fig.~\ref{system}(a) is ideal  when considering the effects of an external magnetic field~\cite{Vidal:2000,Bercioux:2004,Bercioux:2005} or of a spin-orbit coupling~\cite{Bercioux:2004,Bercioux:2005}. If external fields are absent we can consider a stretch version as the one shown in Fig.~\ref{system}(b) that facilitated the identification of lattice symmetries. In case of  homogenous hopping amplitudes, $t_i\equiv t ~\forall i$, each axes perpendicular to the chain and passing though a site H is an inversion axes.

In the following we will focus on the possible set of hopping amplitudes that lead to a dimerisation of the diamond chain --- in the same spirit as in the SSH model~\cite{Su:1979,Su:1980}. Specifically, we have isolated two possible choice for the hopping parameters, the first one  --- named in the following \emph{lattice I} --- is characterised by $t_1=t_4=t$ and $t_2=t_3=t'$; the second one --- named \emph{lattice II} --- is characterised by $t_1=t_3=t$ and $t_2=t_4=t'$. The two lattices are shown in Fig.~\ref{system}(c) and~\ref{system}(d), respectively. From the figure, we immediately observe that contrary to the SSH model, for both dimerisation configurations we cannot find an inversion axes.
These dimerisation configurations are equivalent to consider two SSH chains-sharing a common lattice site --- the H in the specific of our models. The main difference is that in the case of lattice I the two dimerised chains are parallel, whereas in the case of lattice II, the dimerisation of the lower chain is shifted of a unit cell with respect to the upper one. The results we are going to present are strongly affected by the presence of the common lattice site acting as an effective boundary condition.

\subsection{Spectrum and Symmetries}
\subsubsection{Lattice I}
For this lattice, we first write a tight-binding Hamiltonian as,
%
%
\begin{align}\label{hamAreal}
\mathcal{H}_\text{I}  = &\sum_n  t~ h_n^\dag (a_n+b_n) +t'~(a_n^\dag+b_n^\dag)h_{n+1}+ \text{h.c.}\,,
\end{align}
%
%
where $n$ is the index for unit cell. The sub-lattice annihilation(creation) operators are denote by $h_n^{(\dag)}$, $a_n^{(\dag)}$, $b_n^{(\dag)}$ for a particle on the sites H, A and B of unit cell $n$ [Fig.~\ref{system}(c)], respectively. Assuming translational invariance, we can write down the Hamiltonian~\eqref{hamAreal} in the reciprocal space as:
%
%
\begin{subequations}\label{hamAreciprocal}
\begin{eqnarray}\label{hamAreciprocal:1}
[\mathcal{H}_\text{I}]_k &=& d_x(k)\Sigma^\text{I}_x+d_y(k)\Sigma^\text{I}_y, 
\end{eqnarray}
%
%
\noindent with
%
%
\begin{eqnarray}\label{hamAreciprocal:2}
d_x(k)&=&\sqrt{2}(t+t'\cos kL); d_y=\sqrt{2}t'\sin kL\,.
\end{eqnarray}
\end{subequations}
%
%
\noindent In Eq.~\eqref{hamAreciprocal:1} we have introduced the following matrices:
%
\begin{subequations}
\begin{align}\label{sigma1xy}
\Sigma^\text{I}_x =\frac{1}{\sqrt{2}}\begin{pmatrix}
0 &  1  & 0 \\
 1 & 0 & 1 \\
 0 & 1 & 0
\end{pmatrix} &{}&     \Sigma^\text{I}_y = \frac{1}{\sqrt{2}}\begin{pmatrix}
0 &  -\ii  & 0 \\
 \ii & 0 & \ii \\
 0 & -\ii & 0
\end{pmatrix}\,,
\end{align}
%
%
when complemented with the following two:
%
%
\begin{align}\label{sigmazI}
\Sigma^\text{I}_z =\frac{1}{{2}}\begin{pmatrix}
1 &  0  & 1 \\
 0 & -2 & 0 \\
 1 & 0 & 1
\end{pmatrix} &{}& \mathcal{I} = \frac{1}{{2}}\begin{pmatrix}
1 &  0  & 1 \\
 0 & 2 & 0 \\
 1 & 0 & 1
\end{pmatrix}.
\end{align}
\end{subequations}
%
%
We have  that the set $\{\mathcal{I},\Sigma^\text{I}_x,\Sigma^\text{I}_y,\Sigma^\text{I}_z\}$ form an orthogonal basis with SU(2) Lie algebra: $[\Sigma^\text{I}_n,\Sigma^\text{I}_m]=2\ii\epsilon_{nm\ell}\Sigma^\text{I}_\ell$ where  $\epsilon_{nm\ell}$ is Levi-Civita tensor with $n,m,\ell\in \{x,y,z\}$ and $[\cdot,\cdot]$ is the commutator. 
Note that these four matrix operators form a closed subgroup of SU(3). 
As a result, the Hamiltonian Eq.~\eqref{hamAreciprocal} is equivalent to the SSH Hamiltonian \cite{Su:1979,Su:1980}. Subsequently, $\Sigma^\text{I}_z$ (equivalent of $\sigma_z$ operator for SSH model) acts as a chiral operator: $\Sigma^\text{I}_z[\mathcal{H}_\text{I}]_k\Sigma^\text{I}_z=-[\mathcal{H}_\text{I}]_k$. As a result, following SSH model we can expect a topological phase transition changing the ratio $t/t'$.

The Hamiltonian~\eqref{hamAreciprocal} has the following spectrum, 
%
%
\begin{subequations}\label{spectrumA}
\begin{align}
\mathcal{E}^\text{I}_0 &= 0\label{flatA} \\
\mathcal{E}^\text{I}_\pm &= \pm \sqrt{2}\sqrt{t^2+(t')^2+2t't \cos(k L)}
\end{align}
\end{subequations}
%
%
and eigenstates,
%
%
\begin{subequations}\label{vecA}
\begin{align}
\ket{v^\text{I}_0(k)} & = \frac{1}{\sqrt{2}}\left(-1,0,1\right)\label{flatVecA}\\
\ket{v^\text{I}_{\pm}(k)} & = \frac{1}{\sqrt{2}}\left(\frac{\ee^{\ii\phi(k)}}{\sqrt{2}}, \pm 1,\frac{\ee^{\ii\phi(k)}}{\sqrt{2}}\right)\,.
\end{align}
\end{subequations}
%
%
where $\phi(k)= -\arctan\left[\frac{\sin(k L)}{t/t'+\cos(k L)}\right]$. The spectrum consists of two particle-hole dispersive bands with an absolute maximum  and minimum at $k=0$ and relative maximum and minimum at band boundary $k=\pi/L$. The gap at the band boundary is directly proportional to the difference of the two hopping amplitude $m=(t-t')$, and is zero in absence of dimerisation $m=0$. In addition to these two bands, from Eq.~\eqref{flatA} we notice the presence of a flat band at zero energy. This is characterised by an eigenstate~\eqref{flatVecA} that have amplitudes finite and opposite   only of the A and B sites, while it is exactly equal to zero on the H sites.

\subsubsection{Lattice II}
For the lattice II, we redefine the sub-lattice of a unit cell by H, site A on right neighbour of H and site B on the left neighbour of H [Fig.~\ref{system}(d)]. Such redefinition helps us to compare the resulting Hamiltonian from that of lattice I. The Hamiltonian for lattice II then reads:
%
%
\begin{align}\label{hamB}
\mathcal{H}_\text{II}  = \sum_n  t~  ( h_n^\dag a_n&+h_n^\dag b_{n}) +t'~(h_n^\dag a_{n-1}+h_n^\dag b_{n+1}) + \text{h.c.}\,.
\end{align}
This tight-binding Hamiltonian in the reciprocal space reads:
%
%
\begin{eqnarray}\label{hamBreciprocal}
[\mathcal{H}_\text{II}]_k &=& d_x(k)\Sigma^\text{II}_x+d_y(k)\Sigma^\text{II}_y, 
\end{eqnarray}
%
%
with $d_{x,y}(k)$ as in Eq.~\eqref{hamAreciprocal:2} and 
%
%
\begin{align}\label{sigmaIImat}
\Sigma^\text{II}_x =\frac{1}{\sqrt{2}}\begin{pmatrix}
0 &  1  & 0 \\
 1 & 0 & 1 \\
 0 & 1 & 0
\end{pmatrix} &{}&     \Sigma^\text{II}_y = \frac{1}{\sqrt{2}}\begin{pmatrix}
0 &  \ii  & 0 \\
 -\ii & 0 & \ii \\
 0 & -\ii & 0
\end{pmatrix}. 
\end{align}
%
%

%
%
\noindent Comparing Eq.\eqref{sigma1xy} and \eqref{sigmaIImat}, we immediately recognise that
%
%
\begin{subequations}
\begin{align}\label{NoSSH}
\Sigma^\text{I}_x=\Sigma^\text{II}_x \\ 
\Sigma^\text{I}_y\neq\Sigma^\text{II}_y.
\end{align}
\end{subequations}
%
%
The last inequality has an important impact as one can not find an equivalent basis with SU(2) Lie algebra using $\Sigma^\text{II}_{x,y}$ and their commutator. As a result, the lattice II is not equivalent to a SSH model and will not show any topological phase transition by tuning $t/t'$. In order to  look for possible topological transitions in this phase, we first identify the sub-lattice symmetry operator,
%
%
\begin{align}\label{chiralmat}
\Gamma =\begin{pmatrix}
1 &  0  & 0 \\
 0 & -1 & 0 \\
 0 & 0 & 1
\end{pmatrix}.
\end{align}
%
%
This  anti-commutes with $[\mathcal{H}_\text{II}]_k$: $\Gamma^{\dagger}[\mathcal{H}_\text{II}]_k\Gamma =-[\mathcal{H}_\text{II}]_k$. One way to induce topological phase transition in this system is by breaking this sub-lattice symmetry, as we will show later.

The interesting result is that spectrum of lattice II is  identical the one of lattice I:
%
%
\begin{subequations}\label{spectrumB}
\begin{align}
\mathcal{E}^\text{II}_0 &= 0 \\
\mathcal{E}^\text{II}_\pm &= \pm \sqrt{2}\sqrt{t^2+(t')^2+2t't \cos(k L)}
\end{align}
\end{subequations}
%
%
Whereas the eigenstates differs from that of lattice I,
%
%
\begin{subequations}\label{vecB}
\begin{align}
\ket{v^\text{II}_0(k)} & = \frac{1}{\sqrt{2}}\left(-\ee^{\ii \phi(k)},0,\ee^{-\ii \phi(k)}\right) \label{flatBvec}\\
\ket{v^\text{II}_\pm(k)} & = \frac{1}{\sqrt{2}}\left(\frac{1}{\sqrt{2}}\ee^{\ii \phi(k)}, \pm 1,\frac{1}{\sqrt{2}} \ee^{-\ii \phi(k)}\right)\,.
\end{align}
\end{subequations}
%
%
Note that the phase factors enter always with opposite sign on the A and B components. As for lattice I, the flat band eigenstate~\eqref{flatBvec} is characterised by amplitudes finite and opposite on A and B while it is zero on H. This is a consequence on the bipartite nature of the two lattices~\cite{Lin:2015}.
%
%
\begin{figure}[!t]
\begin{center}
\includegraphics[width=\columnwidth]{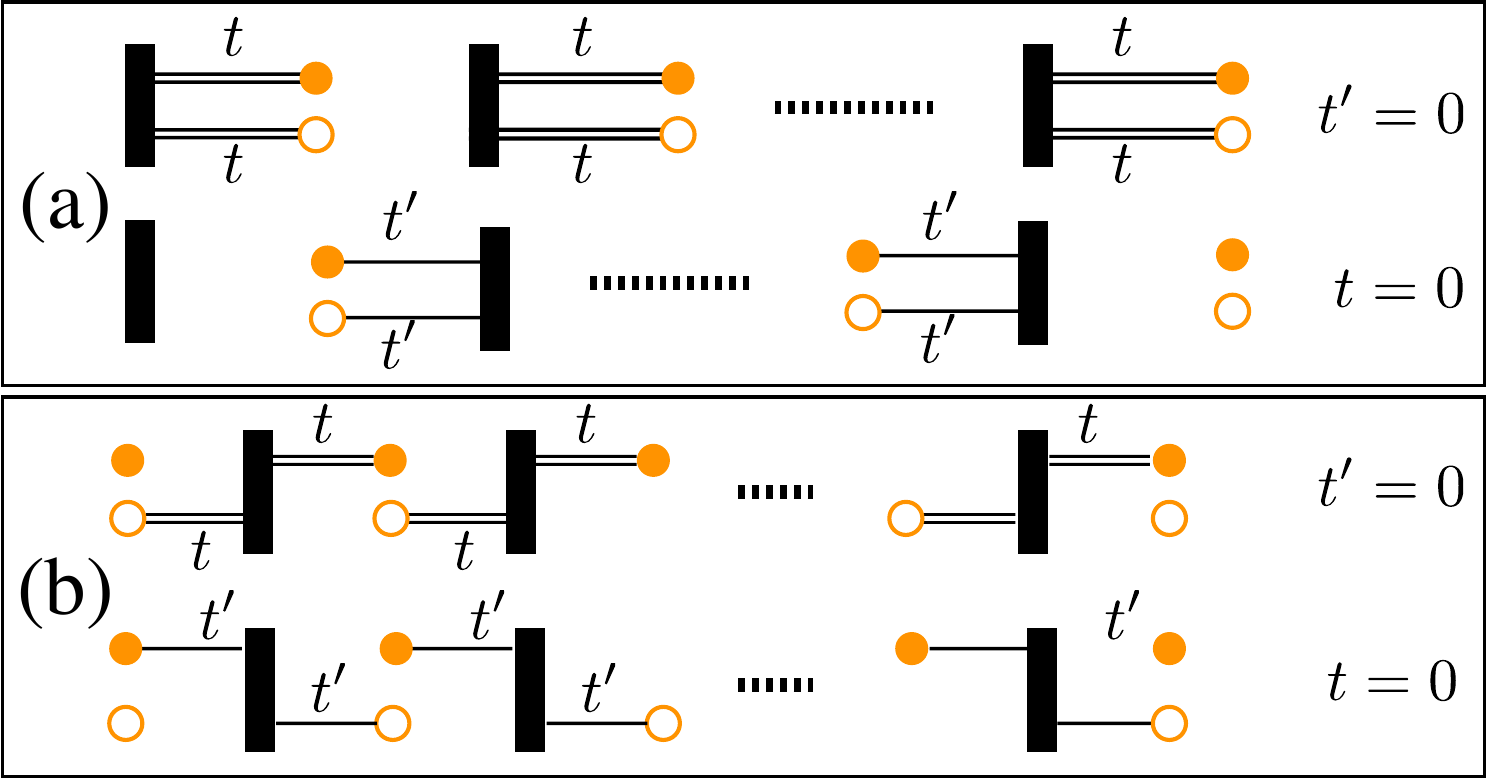}
\caption{\label{dimerised} Fully dimerised version of lattice I (a) and lattice II (b). In the case of lattice I, the two phases cannot be mapped one into the other by any symmetry operation, in the case of lattice II the two phases are equivalent via a $C_2$ rotation.}
\end{center}
\end{figure}
%
%

\section{Topological properties}\label{topo}

In this section we investigate the topological properties of lattice I and II. At first, we look into the fully dimerised case: $t=1$ and $t'=0$ and \emph{vice versa}. As for the SSH model~\cite{Su:1979}, we observe the formation of isolated dimers is a signature of topological properties. The schematic of the two lattices in the fully dimerised phases are shown in two panels of Fig.~\ref{dimerised}. We immediately see that only for the lattice I, we obtain a dimerisation similar to the one of the SSH model~\cite{Asboth:2016}. Whereas for the lattice II, the two fully dimerised cases coincide up to a $C_2$ rotation.

Furthermore, to identify topological phase transition, we have evaluated the Zak phase for the lowest band, which is nothing but the integration of Berry connection over the first Brillouin zone~\cite{ZAK:1989,Asboth:2016}
%
%
\begin{equation}\label{zakphase}
\mathcal{Z}^{a}= \ii \oint dk  \bra{v^a_{-}(k)} \partial_k \ket{v^a_{-}(k)}, ~~a\in\{\text{I,II}\}
\end{equation}
%
%
where the $\ket{v^\text{I,II}_{-}(k)}$ are defined in Eqs.~\eqref{vecA} and~\eqref{vecB}. As expected from our discussion on structure of the Hamiltonians in Eq.~\eqref{hamAreciprocal} and~\eqref{hamBreciprocal}, we find that
%
%
\begin{subequations}
\begin{align}
\mathcal{Z}^\text{I} &= \begin{cases}
 0 & t > t' \\
 \pi & t < t' \\
\end{cases}  \\
\mathcal{Z}^\text{II} &= 0; ~~ \forall t,t'.
\end{align}
\end{subequations}
%
%
This shows that by tuning $t/t'$ for lattice I across the metallic point at $t=t'$, the topological property of lattice I changes. Whereas, there is no topological transition across the metallic point for lattice II.
 
\begin{figure}[!htb]
\begin{center}
\includegraphics[width=\columnwidth]{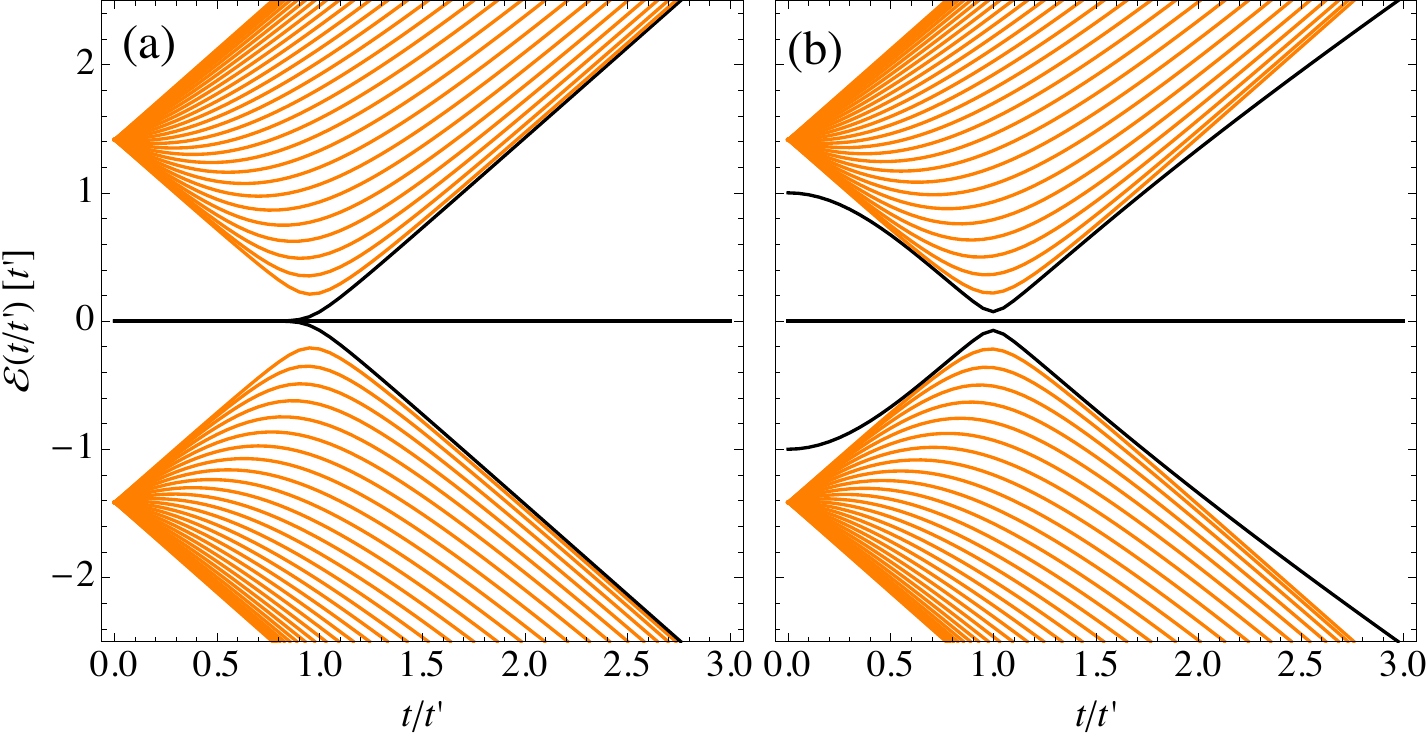}
\caption{\label{Figspec} Energy spectrum for: (a) lattice I and (b) lattice II, as function of $t/t'$ for a finite chain with $N=30$ unit cells.}
\end{center}
\end{figure}

Due to bulk-edge correspondence, the topological properties of lattice I and II can also be seen in appearances of zero energy edge state. This is clearly seen in the spectrum for a finite chain in Fig.~\ref{Figspec}. For lattice I, as one tunes from $t/t' > 1$ to $t/t' < 1$,
%
%
%
%
Fig.~\ref{Figspec}(a) shows that two energy eigenstate from the continuous band transforms to two additional zero energy states. We infer that the wave-functions for these two additional zero energy states corresponds to the left and right edge states of fully dimerised case in Fig.~\ref{dimerised}(a). This also shows the equivalence between lattice I and the SSH model. On the other hand, from spectrum of lattice II in Fig.~\ref{Figspec}(b), we see that there is no crossing as one tunes through the metallic point -- consistent with the absence of a  topological phase transition.

\subsection{Topological charge pumping}

Topological charge pumping refers to quantised current created due to periodic driving of a set of Hamiltonian parameters~\cite{Thouless:1983}. Within the adiabatic limit, such pumping are characterised by appearance of edge states in the spectrum. Such pumping is related to two-dimensional
%
%
\begin{figure}[!htb]
\begin{center}
\includegraphics[width=\columnwidth]{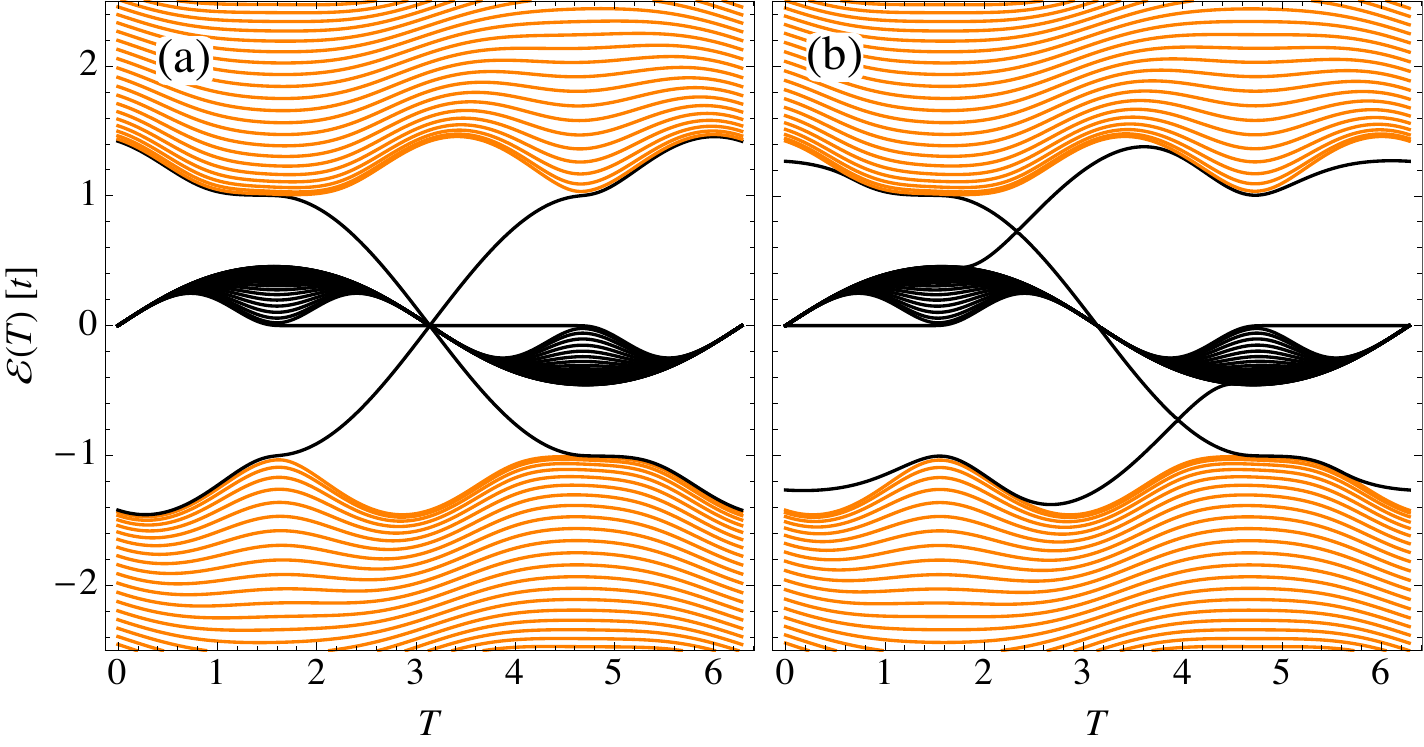}
\caption{\label{FigPump} In the adiabatic limit, the instantaneous spectrum at time T for: (a) lattice I and (b) lattice II, corresponding to the Hamiltonian in Eq.\eqref{hamT}. The dark lines denotes the edge states. In both panels we have considered finite chains with $N=30$ unit cells.}
\end{center}
\end{figure}
%
%
topological insulator and are characterised by Chern numbers. According to Ref.\cite{Brouwer:2011}, for insulators without time-reversal symmetry, a topologically non-trivial pump has a broken sub-lattice symmetry. Moreover, due to adiabaticity condition, at each instance of pumping cycle, the spectrum should be gapped. As a result we choose the following time-dependent Hamiltonian:
%
%
\begin{eqnarray}\label{hamT}
[\mathcal{H}_{a}]_k &=& \sum_{i\in\{x,y\}} d_i(k,T)\Sigma^a_i + \Delta\sin(T) \tau, ~~a\in\{\text{I,II}\}
\end{eqnarray}
%
%
where, 
%
%
\begin{align}\label{Tmat}
\tau = \begin{pmatrix}
1 &  0  & 0 \\
 0 & -1 & 0 \\
 0 & 0 & 0
\end{pmatrix}.
\end{align}
%
%
In Eq.~\eqref{hamT} we have introduced time dependent hopping parameters: $t=1+m\cos(T), t'= 1 - m\cos(T)$, where $T$ is the normalised time and $m$ is the amplitude of tunnelling dimerisation, the Hamiltonian parameters $d_i(k,T)$ now read:
%
%
\begin{subequations}
\begin{align}\label{hamPumping}
d_x(k,T)&=\sqrt{2}\left[\cos\left(\frac{k}{2}\right)^2+m\sin\left(\frac{k}{2}\right)^2\cos(T)\right]; \\
d_y(k,T)&=\frac{1}{\sqrt{2}}\left[1-m\cos(T)\right]\sin kL\,.
\end{align}
\end{subequations}
%
%

It is clear that presence of $\tau$ breaks the chiral/sub-lattice symmetry for lattice I/II. For lattice I, the pumping induces a closed path around the metallic point and acquires a Berry phase due to topological property of time-independent Hamiltonian. As a result, over a cycle, total pumped charge is given by the Chern number of the occupying band (in present case, the lowest band), 
%
%
\begin{eqnarray}
C^a_{-} &=& -\frac{i}{2\pi}\int^{2\pi}_0 dT \int^{\frac{\pi}{L}}_{-\frac{\pi}{L}} dk\left[\partial_k\bra{v^a_{-}(k,T)}\partial_T\ket{v^a_{-}(k,T)} \right. \nonumber\\ 
&-& \left. \partial_T\bra{v^a_{-}(k,T)}\partial_k \ket{v^a_{-}(k, T)}\right],
\end{eqnarray}
%
%
where $\ket{v^a_{-}(k,T)}$ is the instantaneous wave-function of the lower band for Hamiltonian~\eqref{hamT}. We find that for the present model, charge pumping is same as the SSH model, $C^\text{I}_{-}=1$. This is also seen in the instantaneous spectrum of lattice I when considering a finite length chain [Fig.~\ref{FigPump}(a)]. The edge states are denoted by dark lines and at $T=\pi$, where the chiral symmetry is restored, the edge states have zero energy. On the other hand, without pumping, lattice II is topologically trivial insulator. But as seen in Fig.~\ref{FigPump}(b), in the presence of sub-lattice symmetry breaking pumping, edge state arises within a cycle both in the lower and higher band gap. As a result, lattice II has also a topological charge of $C^\text{II}_{-}=1$ and quantised current can be observed. Such topological insulators belong to class A of the classification for adiabatic topological insulators in Ref.\cite{Brouwer:2011}.

\subsection{Continuum limit}
We can analyse the formation of  edge states also studying the zero energy solution of the continuum limit of Hamiltonians~\eqref{hamAreciprocal} and~\eqref{hamBreciprocal}. For both systems the spectra are insensitive to the relative difference of the two hopping amplitudes $m=t-t'$. We can look at an overall mirror symmetry $\Sigma_x^\text{I}$ that exchange the lattice site H with A or B. The expectation value of this symmetry operator at the band boundary is: 
%
%
\begin{equation*}
\langle\Sigma_x^\text{I}\rangle_{k=\frac{\pi}{L}}=\text{sign}[m].
\end{equation*}
%
%
Therefore, a change of sign of $m$ results is a change of parity of the dispersive bands. This change of parity, equivalent to a \emph{band inversion}, is the signature of the existence of zero energy states.

We introduce in the following, for simplicity,  a normalised dimerisation parameter $m$ in the tunnelling as, $t=1+m/2,~t'=1-m/2$. We expand then the Hamiltonians at the band boundary $k=\frac{\pi}{L}+\delta q$. As a result, from \eqref{hamAreciprocal}, \eqref{hamBreciprocal} we get 
%
%
\begin{align*}
d_x(k) &\approx \sqrt{2} m \\
d_y(k) & \approx -\sqrt{2} \delta q\,.
\end{align*}
%
%
Next, we assume that the dimerisation changes as a function of position along chain axis $x$ with $m(x)<0$, $x>0$; $m(x)>0$, $x<0$ , \emph{i.e.}, we places a domain wall at $x=0$. Consequently we use the transformation: $\delta q \rightarrow -\ii \frac{d}{dx}$ to express the Hamiltonian as,
%
%
\begin{equation}\label{SolDirac}
\mathcal{H}_{a} = \sqrt{2}m\Sigma^a_x+\sqrt{2}\left(\ii\frac{d}{dx}\right)\Sigma^{a}_y;~~a\in\{\text{I},\text{II}\}.
\end{equation}
%
%
We then look for the zero energy solution, $\mathcal{H}_{a}\Psi_a(x)=0$ with wave-function $\Psi_a(x)=[\psi^{a}_\text{A}(x),\psi^a_\text{H}(x),\psi^a_\text{B}(x)]^\text{T}$, where A, H, B denote the sub-lattices in a unit cell and T is the transpose operator. Moreover, we impose the constraint of localised solution: $\Psi_a(x\rightarrow \pm \infty)\rightarrow 0$. For lattice I, Eq.~\eqref{SolDirac} we obtain a set Dirac-like equation:
%
%
\begin{subequations}\label{solI}
\begin{align}
&\left(\frac{d}{dx}+m(x)\right)\psi^{\text{I}}_\text{H}(x)= 0 \\
&\left(\frac{d}{dx}-m(x)\right)[\psi^{\text{I}}_\text{A}(x)+\psi^{\text{I}}_\text{B}(x)]= 0.
\end{align} 
\end{subequations}
%
%
For our choice of dimerisation we can find a Jackiw-Rebbi solitonic solution~\cite{Jackiw:1976} of Eqs.~\eqref{solI} that reads: 
%
%
\begin{equation}\label{sol1}
\Psi_\text{I}(x)=\begin{pmatrix}\psi_\text{A} \\ 0 \\ \psi_\text{B}\end{pmatrix} \exp[\int^x_0 m(x')dx']\end{equation}
%
%
On the other hand, for lattice II, the Dirac-like Eq.~\eqref{SolDirac} for the localised solitons are given by,
%
%
\begin{subequations}\label{solII}
\begin{align}
&\left(\frac{d}{dx}-m(x)\right)\psi^{\text{II}}_\text{H}(x)= 0 \label{solIIa}\\
&\left(\frac{d}{dx}+m(x)\right)\psi^{\text{II}}_\text{H}(x)= 0 \label{solIIb}\\
&\left(\frac{d}{dx}+m(x)\right)\psi^{\text{II}}_\text{A}(x)+\left(\frac{d}{dx}-m(x)\right)\psi^{\text{II}}_\text{B}(x)= 0.\label{solIIc}
\end{align}
\end{subequations} 
%
%
We clearly see that, for any shape of dimerisation parameter $m(x)$, the first two equations in~\eqref{solII} and  are incompatible except for $\psi^\text{II}_\text{H}(x)=0$. Moreover, for our choice of dimerisation, the solution of Eq.~\eqref{solIIc} is~\cite{Go:2013},
%
%
\begin{equation}\label{wave_solII}
\psi^\text{II}_\text{B}(x)=\begin{pmatrix}0\\0\\\psi_\text{B}\end{pmatrix}
\exp[\int^x_0 m(x')dx'].
\end{equation}
%
In-spite of the presence of localised solution in lattice II, one can show that this zero energy solution is not protected. To show this, we introduce weak tunnelling between the sites A and B. Then the dynamics of lattice II is governed by the Hamiltonian $\bar{\mathcal{H}}_{a}=\mathcal{H}_{a} + \delta \sum_n (a^\dagger_n b_{n}+b^\dagger_{n} a_{n})$, where the tunnelling amplitude $\delta \ll (t,t')$. Accordingly, in the continuum limit, the equivalent of Eqs.~\eqref{solII} become,
%
%
\begin{subequations}\label{unsolII}
\begin{align}
&\left(\frac{d}{dx}-m(x)\right)\psi^{\text{II}}_\text{H}(x) + \delta\psi^{\text{II}}_\text{B}(x) = -\epsilon \psi^{\text{II}}_\text{A}(x) \label{unsolIIa}\\
&\left(\frac{d}{dx}+m(x)\right)\psi^{\text{II}}_\text{H}(x) - \delta\psi^{\text{II}}_\text{A}(x) = \epsilon \psi^{\text{II}}_\text{B}(x) \label{unsolIIb}\\
&\left(\frac{d}{dx}+m(x)\right)\psi^{\text{II}}_\text{A}(x)+ \left(\frac{d}{dx}-m(x)\right)\psi^{\text{II}}_\text{B}(x)= \epsilon \psi^{\text{II}}_\text{H}(x),\label{unsolIIc}
\end{align}
\end{subequations} 
%
%
where we have introduced a non-zero eigen energy $\epsilon$. By substituting the soliton solution of Eq.~\eqref{wave_solII} in Eqs.~(\ref{unsolIIa}-\ref{unsolIIb}) we get,
$$
\delta \psi^{\text{II}}_\text{B}=0 ; \epsilon \psi^{\text{II}}_\text{B}=0, 
$$
and Eq.~\eqref{unsolIIc} satisfying the solitonic solution. As the A-B tunnelling $\delta \neq 0$, the above equation has solution only for $\psi^{\text{II}}_\text{B}=0$. This shows that, the solitonic solution of Eq. \eqref{wave_solII} is not robust against a small perturbation.  Following the same procedure with the solitonic solution~\eqref{sol1} into the modified Eqs.~\eqref{solI}, shows that this is still a solution of the problem with $\psi_\text{A}=\psi_\text{B}$, thus protected against perturbations.

\section{Discussion, Conclusions and Outlook}

In this manuscript, we have analysed the spectral and the topological properties of two dimerised lattices characterised by the same spectra constituted by two dispersive bands and a flat band in the middle. However, these two configurations have different eigenstates leading to different topological properties. The deepest mathematical difference between lattice I and II is that for the former, the  Hamiltonian can be expressed in terms of \emph{spin}-operators forming a SU(2) subgroup of SU(3). The other dimerisation configuration do not allow to identify a similar sub-group. On the other side, when considering the continuum limit for the case of lattice II, we have show the presence of a zero energy solitonic solution~\cite{Go:2013};  this seems to be an accidental solution and we have shown that is destroyed by any small perturbation applied to the system. By identify the spin operator $\Sigma_z^\text{I}$ as chiral symmetry~\cite{Asboth:2016}, we have that lattice I is in class AIII accordingly to standard Hamiltonian classification whereas lattice II is in class A~\cite{Altland:1997}. Already from this conventional classification scheme we can evince the topological non-trivial nature of lattice I and the trivial nature of lattice II.
%
%
\begin{figure}[t]
\begin{center}
\includegraphics[width=0.8\columnwidth]{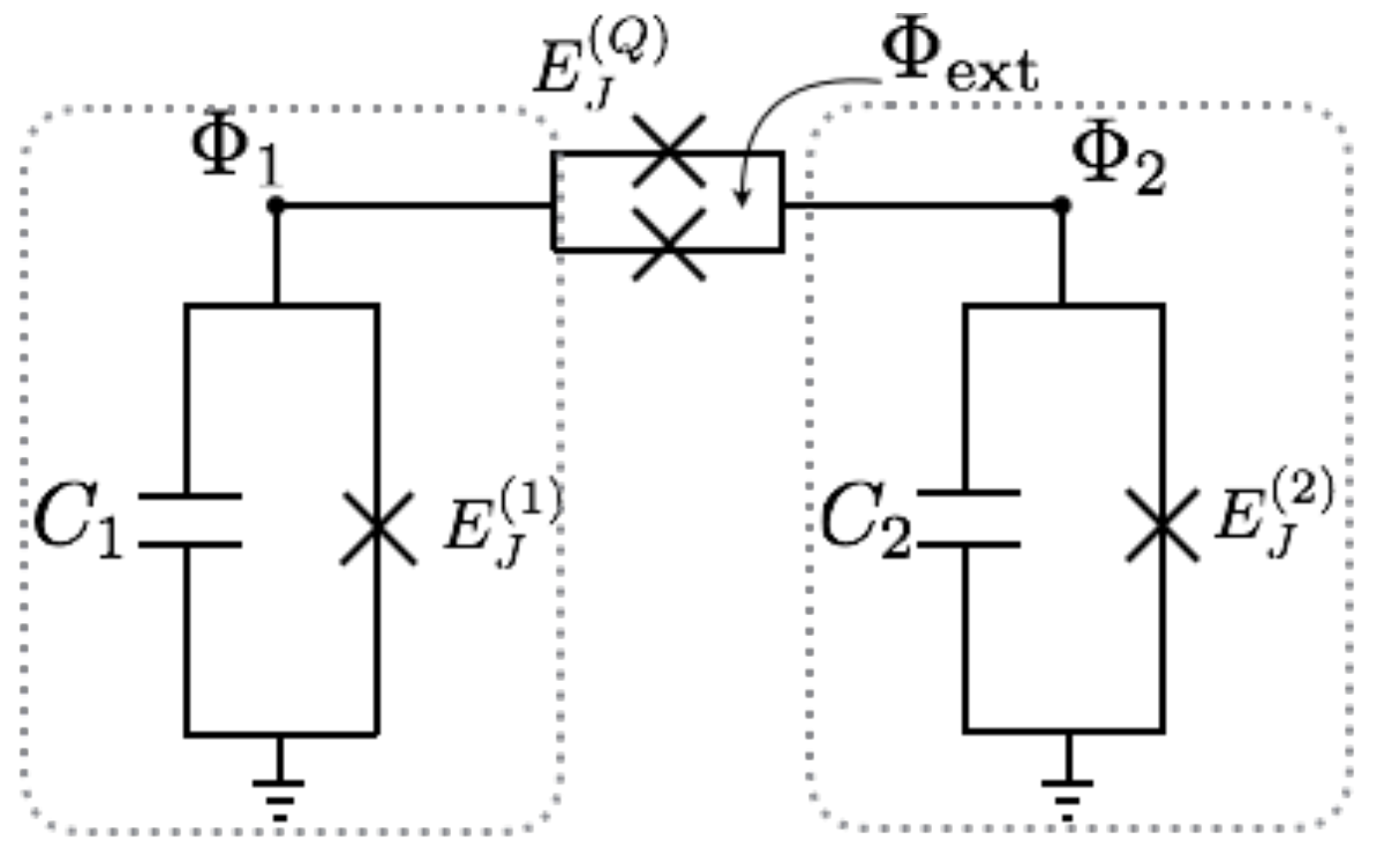}
\caption{\label{implement} Minimal instance where the coupling between two superconducting transmon (dashed boxes) can be tuned at will varying the external classical magnetic flux $\Phi_\text{ext}$ through a SQUID made of two symmetric Josephson junctions in a loop configuration.}
\end{center}
\end{figure}
%
%
Another interesting aspect is the comparison with case of the parallel double SSH model~\cite{Cheon:2015}, where the authors have analysed two parallel SSH chains with different tight-binding coupling between them. This is an effective two-dimensional chain, whereas our model is a quasi-one dimensional one; a direct consequence is that the double SSH chain has a spectrum with four dispersive bands, whereas the diamond chain has three bands where one is always flat. In the presence of adiabatic pumping, the double SSH chain presents edge states very similar to a topological insulator~\cite{Asboth:2016}.  The differences between these double SSH chain and the dimerised diamond chain stem form the different boundary conditions in the chain coupling; in fact, in Ref.~\cite{Cheon:2015} the two chains are coupled by a standard hopping term, whereas in the dimerised diamond chain, the two SSH chains are sharing a lattice site --- the Hub site~\cite{Bercioux:2009,Lin:2015}.

An implementation of the diamond chain with optical lattices has been proposed in Ref.~\cite{Hyrkas:2013}, however this scheme foresees hopping amplitudes identical among all the lattice sites $t_i=t~\forall~i$. Thus, a scheme for implementing the two dimerisation configurations we have proposed needs to exploits some other property, \emph{e.g.},  an internal degree of freedom of the cold atoms loaded in the optical lattice. At the date, quantum simulation of topological insulators has been achieved in several platforms from photonic systems to atomic physics setup. In circuit QED~\cite{Girvin:2011}, a tuneable coupling is obtained with two symmetric Josephson junctions in a SQUID configuration. There, the magnetic flux through the SQUID can activate or deactivate the coupling between two superconducting qubits allowing to explore the desired Hamiltonian in its different phases.

More specifically (see Fig.\ref{implement}), using transmon devices to implement two level systems with local frequency $\omega_{i} = 2e\sqrt{\frac{E_{J}^{(i)}}{C_{i}}}$. Introducing a coupling perturbatively, neglecting the capacitive coupling for the sake of simplicity, then, the potential energy between two transmons is given by: $H_\text{int}=E_{J}^{(Q)} \cos{\left( \frac{\phi_\text{ext}}{\phi_{0}}\right)} \cos{\left( \frac{\phi_{1} - \phi_{2}}{\phi_{0}}\right)}$. Going to a interaction picture with respect to the local on-site Hamiltonians and assuming the rotating-wave approximation, the interacting Hamiltonian is recast into:
%
%
\begin{equation}
H_\text{RW}= - t \left( a_{1}^{\dagger} a_{2} + \text{h.c.} \right)\,,
\end{equation}
%
%
which is exactly of the type we where looking for, where $ t=  \frac{4e^{2} E_{J}^{(Q)} }{2 \sqrt{C_{1} C_{2} E_{J}^{(1)} E_{J}^{(2)} }} \cos{\left( \frac{\phi_\text{ext}}{\phi_{0}}\right)}$ is controlled by the external magnetic flux and with $\frac{\phi_{i}}{\phi_{0}} = \left(\frac{e^{2}}{C_{i}E_{J}^{(i)}}\right)^{1/4} \left( a_{i}^{\dagger} + a_{i} \right)$.

A major issue is to explicitly reveal the topological properties of the system, \emph{i.e.}, by measuring the topological invariants~\cite{Alba2011,Abanin2013,Atala:2013,Bardyn2014,Aidelsburger2014} or observing their edge physics\cite{Hafezi2013}. In this respect, we will show in a forthcoming publication how to reach this goal within circuit QED architectures. 

\begin{acknowledgements}
We acknowledge useful discussions with A. Altland, A. Bernevig, P. Brouwer, R. Resta and I. Souza. The work of DB is supported by Spanish Ministerio de Econom\'ia y Competitividad (MINECO) through the project  FIS2014-55987-P and by the Transnational Common Laboratory \emph{QuantumChemPhys: Theoretical Chemistry and Physics at the Quantum Scale}. The work of ER is supported by a UPV/EHU grant EHUA15/17 and by Spanish Ministerio de Econom\'ia y Competitividad through the project MINECO/FEDER FIS2015-69983-P. 
\end{acknowledgements}
\bibliography{Biblio}
\end{document}